\documentclass[a4paper,12pt]{article}\pdfoutput=1
\usepackage{graphicx, rotating, slashed}
\ifx\pdfoutput\undefined
\usepackage[dvips,bookmarks]{hyperref}	
\else
\usepackage{hyperref}	
\fi
\hypersetup{colorlinks,bookmarksopen,bookmarksnumbered,citecolor=verdes,
linkcolor=blus,pdfstartview=FitH,urlcolor=rossos}
\def\hhref#1{\href{http://arxiv.org/abs/#1}{#1}} 

\newcommand{\fig}[1]{~\ref{fig:#1}}
\usepackage{multicol}
\usepackage{color}
\definecolor{rosso}{cmyk}{0,1,1,0.4}
\definecolor{rossos}{cmyk}{0,1,1,0.55}
\definecolor{rossoc}{cmyk}{0,1,1,0.2}
\definecolor{blu}{cmyk}{1,1,0,0.3}
\definecolor{blus}{cmyk}{1,1,0,0.6}
\definecolor{bluc}{cmyk}{1,1,0,0.1}
\definecolor{verde}{cmyk}{0.92,0,0.59,0.25}
\definecolor{verdec}{cmyk}{0.92,0,0.59,0.15}
\definecolor{verdes}{cmyk}{0.92,0,0.59,0.4}

\newcommand{\eq}[1]{~{\rm (\ref{eq:#1})}}

\newcommand{\MeV}{\,{\rm MeV}}
\newcommand{\GeV}{\,{\rm GeV}}
\newcommand{\GeVn}{\,{\rm GeV/nuc}}
\newcommand{\TeV}{\,{\rm TeV}}
\newcommand{\cm}{\,{\rm cm}}
\newcommand{\sr}{\,{\rm sr}}

\def\circa#1{\,\raise.3ex\hbox{$#1$\kern-.75em\lower1ex\hbox{$\sim$}}\,}

\newcommand{\NP}{Nucl. Phys.}

\newcommand{\PR}{Phys. Rev.}

\newcommand{\beq}{\begin{equation}}
\newcommand{\eeq}{\end{equation}}

\newcommand{\bea}{\begin{eqnarray}}
\newcommand{\eea}{\end{eqnarray}}

\font\tenrsfs=rsfs10 at 12pt
\font\sevenrsfs=rsfs7
\font\fiversfs=rsfs5
\newfam\rsfsfam
\textfont\rsfsfam=\tenrsfs
\scriptfont\rsfsfam=\sevenrsfs
\scriptscriptfont\rsfsfam=\fiversfs
\def\mathscr#1{{\fam\rsfsfam\relax#1}}

\def\circa#1{\,\raise.3ex\hbox{$#1$\kern-.75em\lower1ex\hbox{$\sim$}}\,}
\makeatletter

\def\art{\@ifnextchar[{\eart}{\oart}}
\def\eart[#1]#2#3#4#5#6{{\rm #2}, {#3 #4} {\rm (#6) #5} [arXiv:{\hhref{#1}}]}
\newcommand{\oart}[5]{{\rm #1}, {#2 #3} {\rm (#5) #4}}
\def\hepart[#1]#2{{\rm #2, arXiv:\hhref{#1}}}
\newcounter{alphaequation}[equation]
\def\thealphaequation{\theequation\hbox to
0.6em{\hfil\alph{alphaequation}\hfil}}
\def\eqnsystem#1{
\def\@eqnnum{{\rm (\thealphaequation)}}
\def\@@eqncr{\let\@tempa\relax \ifcase\@eqcnt \def\@tempa{& & &} \or
 \def\@tempa{& &}\or \def\@tempa{&}\fi\@tempa
 \if@eqnsw\@eqnnum\refstepcounter{alphaequation}\fi
\global\@eqnswtrue\global\@eqcnt=0\cr}
\refstepcounter{equation} \let\@currentlabel\theequation \def\@tempb{#1}
\ifx\@tempb\empty\else\label{#1}\fi
\refstepcounter{alphaequation}
\let\@currentlabel\thealphaequation
\global\@eqnswtrue\global\@eqcnt=0 \tabskip\@centering\let\\=\@eqncr
$$\halign to \displaywidth\bgroup \@eqnsel\hskip\@centering
$\displaystyle\tabskip\z@{##}$&\global\@eqcnt\@ne
\hskip2\arraycolsep\hfil${##}$\hfil& \global\@eqcnt\tw@\hskip2\arraycolsep
$\displaystyle\tabskip\z@{##}$\hfil
\tabskip\@centering&\llap{##}\tabskip\z@\cr}
\def\endeqnsystem{\@@eqncr\egroup$$\global\@ignoretrue} \makeatother

\begin{document}

\begin{center}
 CERN-PH-TH/2009-149

\bigskip\bigskip
\bigskip

{\LARGE\bf\color{magenta}
Enhanced anti-deuteron Dark Matter  signal \\[5mm]
and  the implications of  PAMELA}  \\[5mm]

\medskip
\bigskip\color{black}\vspace{0.6cm}
{
{\large\bf Mario Kadastik$^a$, Martti Raidal$^a$, Alessandro Strumia$^{bc}$}
}
\\[7mm]
{\it $^a$NICPB, Ravala 10, 10143 Tallinn, Estonia} \\
{\it $^b$Dipartimento di Fisica dell'Universit{\`a} di Pisa and INFN, Italia}\\
{\it $^c$ CERN, PH-TH, CH-1211, Geneva 23, Suisse}
\bigskip\bigskip\bigskip

{\large
\centerline{\large\bf Abstract}

\begin{quote}

We show that the jet structure of  DM annihilation or decay products enhances the  $\bar d$ production 
rate by orders of magnitude compared to the previous computations done assuming a spherically symmetric coalescence model.
In particular,  in the limit of heavy DM, $M\gg m_p$, we get a constant rather than 
$1/M^2$ suppressed $\bar d$ production rate. 
Therefore, a detectable $\bar d$ signal is compatible with the lack of an excess in the $\bar p$ PAMELA flux.
Most importantly,  cosmic $\bar d$ searches become sensitive to the annihilations or decays of heavy DM, 
suggesting to extend the experimental  $\bar d$ searches  above the ${\cal O}(1)$~GeV scale.

\end{quote}}

\end{center}
\newpage

\section{Introduction}

The cosmological DM abundance is naturally produced in thermal freeze-out if Dark Matter (DM) 
has weak interactions  and a TeV-scale mass $M$, that in appropriate models can be lowered
down to the weak scale, 100 GeV.
This scenario can be tested searching for  DM annihilation (or decay) products in cosmic rays.
In view of astrophysical backgrounds, a particularly sensitive signal is an
excess in cosmic-ray anti-particles:  positrons, anti-protons $\bar p$ and anti-deuterium $\bar d$.  
According to the coalescence prescription~\cite{Csernai}, a $\bar d$ is formed when DM produces 
a $\bar p$ and a $\bar n$ with momentum difference below $p_0 \approx 160 \MeV$.
The standard formula for the $\bar d$ spectrum, obtained under the assumption of {\em spherical symmetry} of the events, 
 in terms of the anti-nucleon 
($\bar p$ plus $\bar n$) 
energy spectrum per annihilation, $dN_N/dT$, is~\cite{d-bar,d-bar2, Fornengo, Cirelli, Ibarra}
\beq 
\frac{dN_{\bar d}}{dT_{\bar d}} = \frac{p_0^3}{3k_{\bar d}m_p}\left(\frac{1}{2}\frac{dN_N}{dT}\right)^2_{T=T_d/2},
\label{eq:eq1}
\eeq
where  the kinetic energies $T=E-m$ are    $T_p = T_n = T_d/2,$  $m_p = m_n=m_d/2$ and
 $k_{\bar d}=\sqrt{T_{\bar d}^2+2m_d T_{\bar d}}$.
Eq.\eq{eq1} implies a $\bar d$ yield suppressed by $1/M^2$ for large $M$.  
This result is qualitatively wrong. 
Increasing $M$ just increases the boost of the primary DM annihilation products, 
giving rise, due to Lorentz symmetry, to an essentially constant $\bar d$ production rate
 with energy roughly proportional to $M$. The reason for this fundamental discrepancy is caused by the fact that
the spherical approximation misses the jet structure of the DM annihilation products.

In this letter we show that for  $M\gg m_p$ the angular proximity of the produced $\bar p,\bar n$ enhances 
the $\bar d$ yield, possibly by orders of magnitude. We critically compare the standard spherical approximation results
with our Monte Carlo approach to $\bar d$ production, presenting the $\bar d$ energy spectra for the various  
DM annihilation or decay channels into $W^+ W^-, ZZ,q\bar{q}, b\bar{b},t\bar{t},hh$ and comment on the
astrophysical $\bar d$ background produced mostly in cosmic ray $pp$ and $p\bar p$ collisions.
We propagate $\bar d$ in the Milky Way, studying the phenomenology and the prospects for DM produced
$\bar d$ searches at AMS-2, in the light of the PAMELA $\bar p$ observations.
We find that the $\bar d$ signal below 1~GeV is strongly enhanced increasing the chances of $\bar d$ detection at AMS-2 
even for the standard thermal DM annihilation cross section. This result is consistent with the lack of $\bar p/p$ excess
in PAMELA. Due to the qualitatively different large $M$ behavior of the production rate, 
our result drastically enhance the $\bar d$ production  at high energies.
Therefore  the cosmic ray $\bar d$ flux produced in heavy DM annihilations or decays 
exceeds the estimated background, and AMS-2 and future $\bar d$ experiments 
become sensitive to DM if they extend their sensitivity to $\bar d$ above 1~GeV.

\section{Spherical-cow vs Monte Carlo}\label{cow}
$\bar d$ is formed via $\bar p \bar n\to \bar d\gamma$, that has a large cross-section due to the small binding energy of $\bar d$, which therefore has a
spatially extended wave-function or equivalently a strongly peaked
wavefunction $\psi(\Delta k) \equiv \langle \bar d|\bar p \bar n \rangle$  in momentum space.
Here $\bar d$ has momentum $k_d$ and energy $E_d$
 and $\Delta k$ is the relativistically invariant
relative momentum between $\bar p$ (with momentum $\vec k_p$ and energy $E_p$) and $\bar n$
(with momentum $\vec k_n$ and energy $E_n$):
\beq\label{eq:Deltak} \Delta k^2 = |\vec k_p - \vec k_n|^2 - (E_p -E_n)^2 + (m_n-m_p)^2 , \eeq
where we can neglect $m_p - m_n= 1.29\MeV $.
The amplitude for $\bar d$ production in DM annihilations,
${\rm DM~DM}\to\bar d$, can be computed as
\beq   \langle \bar d |{\rm DM~DM}\rangle =
\sum_{\bar p,\bar n} \langle \bar d|\bar p \bar n \rangle
 \langle \bar p\bar n|{\rm DM~DM}\rangle ,
\eeq
giving the `coalescence approximation'~\cite{Csernai}:
the probability $|\langle \bar d|\bar p \bar n \rangle|^2$ that a $\bar p$ and a $\bar n$ coalesce to form a $\bar d~$ 
is approximated as a narrow step function  $\Theta(\Delta k - p_0)$, that drops from unity to zero if $\Delta k$  is larger than $p_0$.
Here $p_0$ is a constant (to be extracted from data later) 
that can be estimated as $p_0 \sim \sqrt{ m_d B_d}\sim 60\MeV$ assuming that 
$\bar d$ production happens until the relative $\bar p,\bar n$ kinetic energy is smaller than
the deuteron binding energy $B_d = 2.2\MeV$.
The total $\bar d$ yield therefore is
\beq\label{eq:Nd}
N_d =  \int  dN_p ~dN_n \Theta(\Delta k^2 - p_0^2)=
 \int d^3k_p \, d^3 k_n \frac{dN_{p}dN_n}{d^3 k_p\,d^3 k_n}
 \Theta(\Delta k^2 - p_0^2).\eeq
In the non-relativistic limit $k_{p,n}\ll m_{p,n}$ and for small $p_0$
the region that satisfies $\Delta k < p_0$ at fixed $\vec k_n$ is a sphere in $\vec k_p$
centered on $\vec k_n$ with radius $p_0$ and volume $4\pi p_0^3/3$.
In general the sphere gets dilatated along the direction $\vec k_{p}\approx \vec k_n$ by a relativistic Lorentz
factor $\gamma_{p}\approx \gamma_n\approx \gamma_d$.
Multiplying eq.\eq{Nd} times $1 = \int d^3k_d \,\delta(\vec k_d - \vec k_p - \vec k_n)$ we finally get,
in the limit of small $ p_0\ll M/\gamma_{p,n}$, the $\bar d$ momentum distribution:
\beq \label{eq:dNd}
\gamma_d \frac{dN_d}{d^3 k_d} = \frac{1}{8} \frac{4\pi p_0^3}{3} \gamma_n  \gamma_p  \frac{dN_{p}dN_n}{d^3 k_p\,d^3 k_n},\eeq
where $\vec k_p = \vec k_n = \vec k_d/2$.\footnote{The extra factor of 8 with respect to the
equation used in papers \cite{d-bar}-\cite{Ibarra} comes from $d^3 k_d = 8 \,d^3 k_{p,n}$.  In the final result this difference gets
compensated by a value of $p_0$ twice larger than the one adopted in those papers.
In our Monte Carlo computation of the coalescence condition $\Delta k < p_0$ it is important that
we fix factors of 2 so that our $p_0$ really is the radius of the coalescence sphere.
}
Eq.\eq{dNd}  is relativistically invariant as it contains the usual relativistic phase space 
$d^3k/2E = d^4k~\delta(E^2-k^2-m^2)$.  
\subsection*{The spherical approximation}
Previous computations proceed assuming spherical symmetry, $d^3k = 4\pi k^2~dk$, 
and uncorrelated $\bar p$, $\bar n$ distributions:
\beq 
 \frac{dN_{p}dN_n}{d^3 k_p\,d^3 k_n}= \frac{dN_{p}}{d^3 k_p}\cdot
  \frac{dN_n}{d^3 k_n}
  \qquad\hbox{implying}\qquad
  \frac{E}{m}\frac{dN}{d^3k} = \frac{1}{4\pi k m}\frac{dN}{dE}.
  \eeq
Writing the result in terms of the adimensional  $x_i = T_i/M$ (so that $0\le x_{d,p,n}<1$
and $x_p=x_n=x_d/2$) one gets eq.\eq{eq1} i.e.
 \beq \label{eq:spher}
  \frac{dN_d}{dx_d} = \frac{p_0^3}{3M^2m_p} \frac{1}{\sqrt{x_d^2 +4m_p x_d/M}}\frac{dN_p}{dx_p} \frac{dN_n}{dx_n},\eeq
 which is explicitly suppressed by $1/M^2$ for large DM mass $M$.

{\em This is qualitatively wrong}.
Consider for example the ${\rm DM~DM}\to W^+ W^-$ annihilation mode. Increasing $M$ increases the boost of each $W$, and thereby the boost of the
anti-deuterons from $W$ decay, but the $\bar d$ number stays fixed.  
Neglecting QED final state radiation (FSR),  for $M\gg M_W$
one should get a constant, $M$-independent function for $dN_d/dx_d.$
Obviously the problem is in the `spherical cow' approximation~\cite{cow}. Due to the $W^\pm$ boost the events are highly non spherical
and SM particles are concentrated in two back-to-back jets, enhancing the probability of having  $\bar p\bar n$ pairs with 
small momentum difference $\Delta k < p_0$.
A similar argument applies to DM annihilations or decays into colored particles, such as $q\bar{q}$.
Hadronization leads to QCD jets, rather than to spherical events.
Thereby the spherical approximation can grossly underestimate the $\bar d$ production.

Going to less relevant aspects that control order one factors, the 
analytic spherical approximation can also over-estimate the $\bar d$ yield, by neglecting
anti-correlations between $\bar n$ and $\bar p$ or the fact that no $\bar d$ is obtained
if only one anti-nucleon is present per event.
As an example, we consider again the $W^+W^-$ mode: within the spherical approximation
a $\bar d$ can form coalescing a $\bar p$ from $W^-$
with a $\bar n$ from $W^+$, but this process is highly suppressed because the $W^+$ and $W^-$ go back to back.


\begin{figure}[t]
\begin{center}
\includegraphics[width=\textwidth]{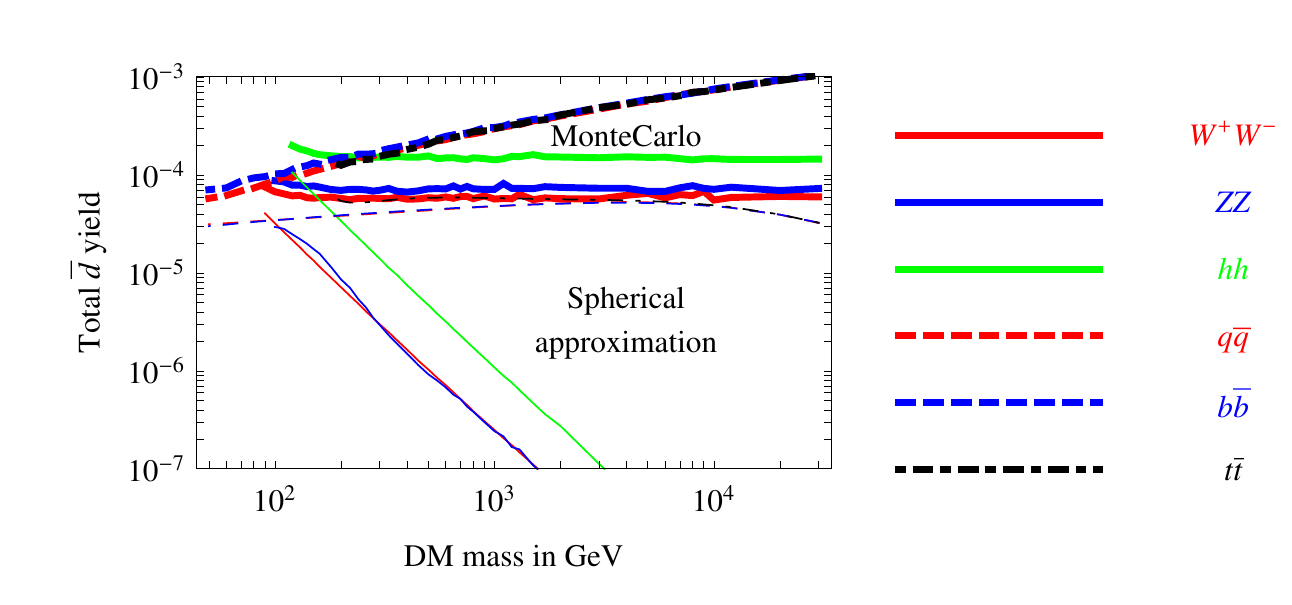}
\caption{\em\label{fig:dyield}  { Total $\bar d$ yield} per DM annihilation
as a function of the DM mass.
The thick upper lines present the Monte Carlo result, and the lower thin lines are the spherical approximation.
The annihilation modes are into $W^+ W^-$ (red), $ZZ$ (blue), $hh$ (green)
$t\bar t$ (black dot-dashed), $b\bar b$ (blue dashed), light quarks $q\bar q$ (red dashed).
}
\end{center}
\end{figure}

\begin{figure}[t]
\begin{center}
\includegraphics[width=0.95\textwidth]{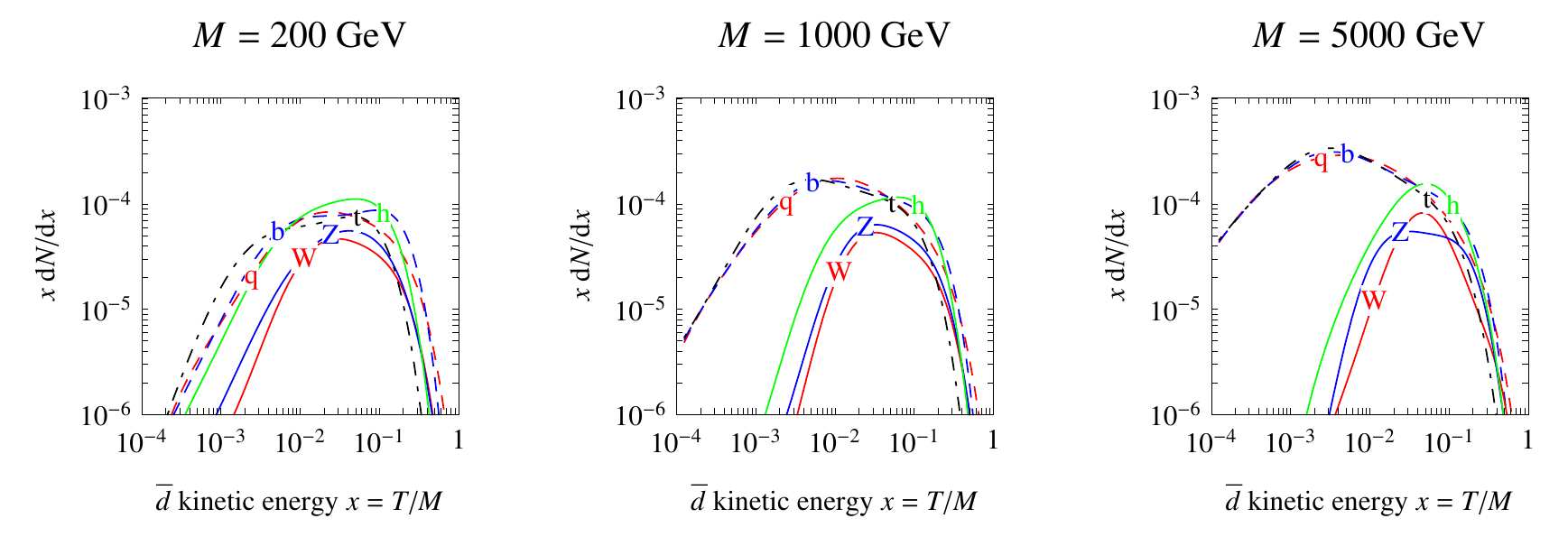}
\caption{\em\label{fig:dNdxM}  { Monte Carlo results for
the $\bar d$ spectra} $x \cdot dN/dx$ produced per DM annihilation into $W^+ W^-$, $ZZ$, $hh$
$t\bar t$, $b\bar b$, $q\bar q$.
We assumed the DM mass $M=200\GeV$ ($1\TeV$) [$5\TeV$] in the left (middle) [right] panel.
The notation is the same as in fig.\fig{dyield}.
}
\end{center}
\end{figure}

\subsection*{The Monte Carlo approach}
In order to take into account the jet structure of the events and  the correlations between the $\bar p,\bar n$ momenta
we compute the $\bar d$ spectrum   by searching {\em event-by-event} for the $\bar n,\bar p$ pair(s) which have
relativistically invariant momentum difference $\Delta k$ smaller than $p_0$.
We verified that the spherical uncorrelated approximation of eq.\eq{spher} is reproduced if we first merge many events, and later
coalesce $\bar p$ with $\bar n$ without imposing that they come from the same event.

Various experiments extracted compatible values of
$p_0$ from data about $\bar d$ production in hadronic and $e^+e^-$ collisions.
Presumably these studies adopted the `spherical cow' approximation rather than
performing a Monte Carlo computation.  Giving the relatively low energies involved this should not make
a large difference; anyhow we here prefer to directly extract $p_0$ from the ALEPH data~\cite{ALEPH}:
one hadronic $Z$ decay at rest gives rise to $(5.9 \pm 1.9)~10^{-6}$ $\bar d$ in the momentum range
$0.62\GeV< k_d < 1.03\GeV$ and angular range $|\cos\theta|<0.95$.
According to our Monte Carlo computation, this translates into $p_0 = 162 \pm 17\MeV$.
Should $p_0$ have a value different from the $p_0=160\MeV$ adopted here
for both the DM signal and the astrophysical background (as computed in~\cite{d-bar,Fornengo})
the  $\bar d$ energy spectra get rescaled roughly by an overall $(p_0/160\MeV)^3$ factor.


We performed a Monte Carlo study by generating a huge number of events (up to $10^7$ per DM DM annihilation,  
and up to $10^9$ events when studying $pp$ and $p\bar p$ collisions) with {\sc Pythia} 8~\cite{Pythia}, 
directly implementing the condition $\Delta k < p_0 = 160\MeV$ for $\bar d$ production.
Such computing-power demanding results have been obtained using the EU Baltic Grid facilities~\cite{BG}.

Fig.\fig{dyield} shows the total number of $\bar d$ produced per DM annihilation as
function of the DM mass for various annihilation modes, comparing our Monte Carlo result
with the spherical approximation, which can under-estimate the $\bar d$ yield by various orders of magnitude.
The same $p_0 = 160\MeV$ is assumed in both cases. 

\begin{figure}[t]
\begin{center}
\includegraphics[width=0.95\textwidth]{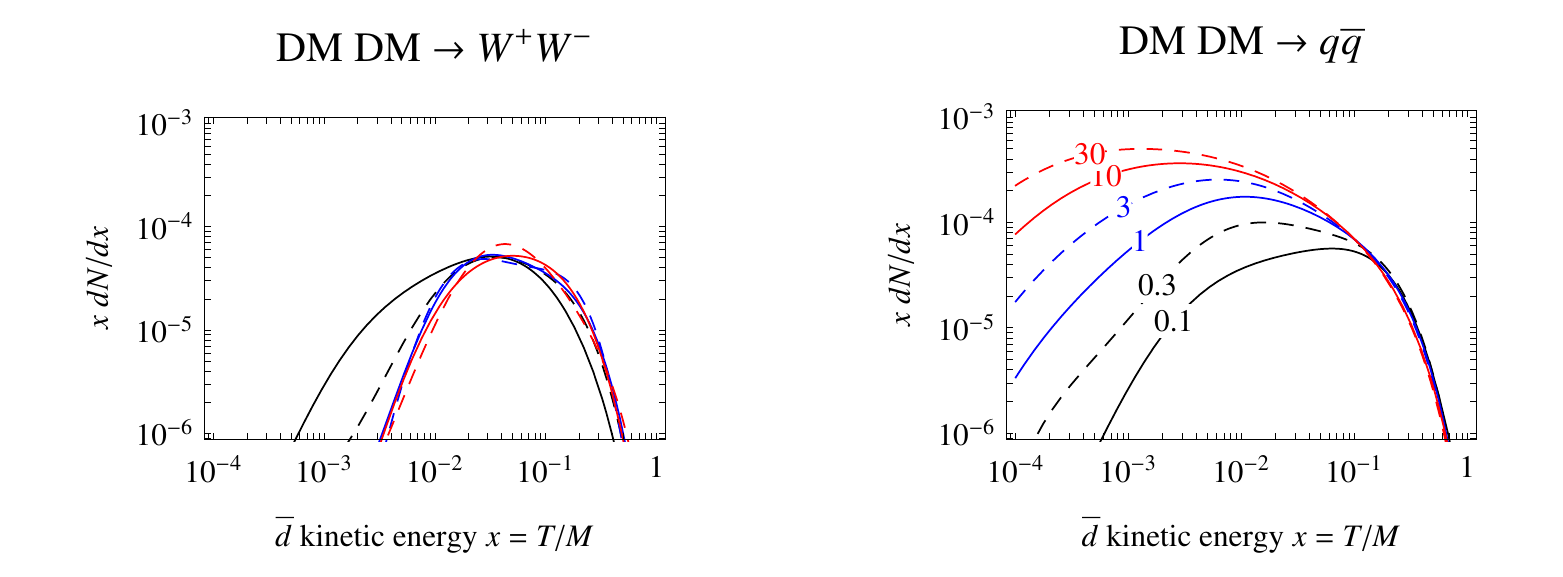}
\caption{\em\label{fig:dNdx} 
{ Monte Carlo results for the $\bar d$ spectra} per DM annihilation, $x\cdot dN/dx$ for $p_0 = 160\MeV$.
We consider DM masses $M=0.1,1,10\TeV$ (black, blue, red continuous curves)
and $M=0.3,3,30\TeV$ (black, blue, red dashed curves)
and the indicated DM annihilation modes.
}
\end{center}
\end{figure}

Fig.\fig{dNdxM} shows our Monte Carlo results for
the $\bar d$ spectra computed for
three values of the annihilating DM mass $M$.
The same spectra also hold for decaying DM, after replacing $M_{\rm ann} \to M_{\rm dec}/2$.
As we expected, the result has only a minor dependence on $M$
and is thereby qualitatively different from the `spherical-cow' approximation that would give a $1/M^2$ suppression.
There are three classes of qualitatively different cases: DM annihilations i) into $W,Z,h$
(we assume a Higgs mass $m_h = 120\GeV$); ii) into quarks $q,b,t$ or iii) into leptons.
The latter case gives no $\bar d$.
To compare the former two cases that give $\bar d$, we focus on i)
DM DM $\to W^+W^-$ and ii) DM DM $\to q \bar q$, and show the $\bar d$ spectra  in fig.\fig{dNdx}a and b, respectively,  
for various values of the DM mass $M$.
In the $W^+ W^-$ case the $\bar d$ spectrum only mildly depends on the DM mass.
Neglecting FSR, all $\bar d$ should have  $x > m_d/M_W = 0.05$; the
small $\bar d$ flux at smaller $x$ is due to electroweak FSR.
In the $q\bar q$ case the $\bar d$ spectrum at smaller $x$ increases with $M$ rather than being suppressed as $1/M^2$.
This is due to QCD FSR that roughly scales as $\alpha_3 \ln (M/m_p)$.

The Monte Carlo results differ both
qualitatively and quantitatively
from the previous studies  of the $\bar d$ spectra in DM annihilations or decays. 
To draw conclusions about the detectability of the signal, we also need to study possible changes in the 
 astrophysical $\bar d$ background, mainly generated by collisions of cosmic-ray $p$ with energy $E_p$ on $p$ at rest.
In view of the kinematical threshold for $\bar d$ production ($E_p\circa{>}30\GeV$) and
of the energy spectrum of  cosmic protons (roughly proportional to $E_p^{-3}$),
$\bar d$ production is dominated by $E_p \sim 60\GeV$,
in the range of validity of the parton model in {\sc Pythia}.
Our semi-quantitative results for the $\bar d$  background suggest 
a reasonable agreement with the spectra of~\cite{d-bar,Fornengo}. This is an expected result because 
the center of mass energy in cosmic $pp$ collisions is small and, in this case, the 
uncorrelated spherical approximation
is expected to work reasonably well. 
However this issue needs to be precisely investigated.

\medskip

Some remarks are in order.
First, we computed $p,n$ allowing all other hadrons to decay despite that the life-time of
some strange baryons, such as the $\Xi=uss$, is longer than the size of deuterium.
This effect should already have been taken into account when extracting the value of $p_0$ from high-energy
experimental data from its definition of eq.\eq{dNd}. 
Second,  DM in general annihilates into various primary channels $k$.
According to eq.\eq{eq1} one should sum their contributions to the $\bar p,\bar n$ spectra rather than to
the $\bar d$ spectrum, getting $(\sum_k dN^{(k)}/dx)^2 \neq \sum_k (dN^{(k)}/dx)^2$.
Our Monte Carlo  result instead amounts to sum incoherently over all
primary annihilation channels $k$ as well as all  secondary and  tertiary  contributions in the decay chain.

\begin{figure}[t]
\begin{center}
\includegraphics[width=0.45\textwidth]{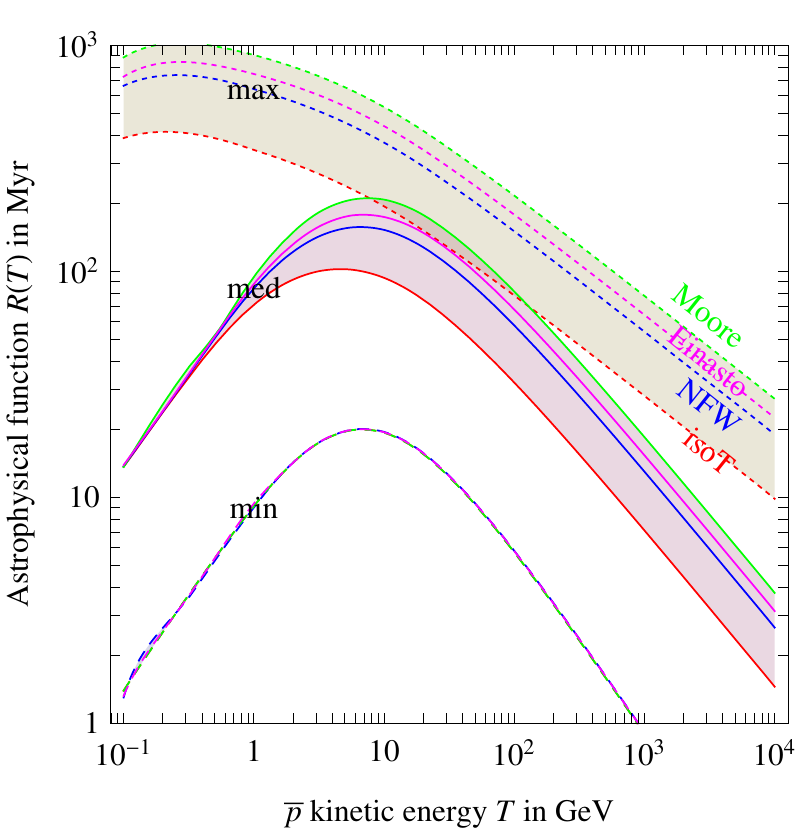}\qquad
\includegraphics[width=0.45\textwidth]{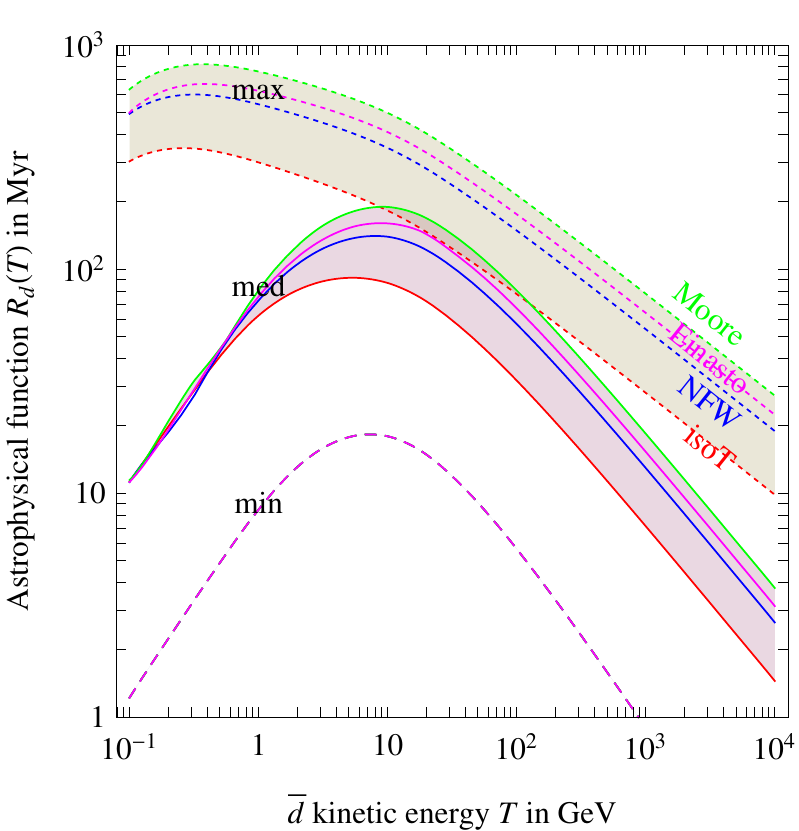}
\caption{\em\label{fig:HaloI} 
The $\bar p$ (left) and $\bar d$ (right) astrophysical functions, $R(T)$ and $R_d(T)$ of eq.\eq{RT}, computed under different assumptions.
In both cases, the dashed (solid) [dotted] bands assumes the {\rm min (med) [max]} propagation configurations.
Each  band contains 4 lines, that correspond to the isothermal (red lower lines), NFW (blue middle lines), Einasto (magenta) and Moore (green upper lines) DM density profiles.
}
\end{center}
\end{figure}

\section{Cosmological fluxes}\label{results}

To compute the $\bar d$ flux in the solar system we consider 
four possible  Milky Way DM density profiles $\rho(r)$~\cite{rho(r)}: 
\beq \frac{\rho(r)}{\rho_\odot} = \left\{\begin{array}{ll}
(1+r_\odot^2/r_s^2)/(1+r^2/r_s^2) & \hbox{isothermal, $r_s = 5\,{\rm kpc}$}\\
(r_\odot/r)(1+r_\odot/r_s)^2/(1+r/r_s)^2 & \hbox{NFW, $r_s = 20\,{\rm kpc}$}\\
(r_\odot/r)^{1.16}(1+r_\odot/r_s)^2/(1+r/r_s)^{1.84} & \hbox{Moore, $r_s = 30\,{\rm kpc}$}\\
\exp(-2[(r/r_s)^\alpha - (r_\odot/r_s)^\alpha]/\alpha)& \hbox{Einasto, $r_s = 20\,{\rm kpc}$, $\alpha=0.17,$}\\
\end{array}
\right.
\eeq
keeping fixed the local DM density $\rho(r=r_\odot) = \rho_\odot=0.3\GeV/\cm^3$.
Concerning diffusion of charged $\bar d$ in the galaxy, we 
approximate the diffusion region as a cylinder with height $2L$ centered on the galactic plane, a
constant diffusion coefficient $K = K_0 E^\delta$
and a constant convective wind directed outward perpendicularly to the galactic plane.
We consider the min, med, max propagation models~\cite{minmedmax} for $\bar p,\bar d$, which are characterized
by the following astrophysical parameters,
\beq \begin{tabular}{ccccc}
Model  & $\delta$ & $K_0$ in kpc$^2$/Myr & $L$ in kpc & $V_{\rm conv}$ in km/s \\
\hline 
min  & 0.85 &  0.0016 & 1  &13.5\\
med  & 0.70 &  0.0112 & 4  & 12 \\
max  & 0.46 &  0.0765 & 15 &5
\end{tabular}\label{eq:proparam}\quad . \eeq
Finally, one must take into account annihilations of $\bar d$ on interstellar protons and Helium in the galactic plane
(with a thickness of $h=0.1\,{\rm kpc} \ll L$) with rate $\Gamma_{\rm ann} $~\cite{d-bar}.
The solution to the diffusion equation for the  energy spectrum of the $\bar d$ number density, $f$,
\beq 
\label{eq:diffeqp}
- K(T)\cdot \nabla^2f + \frac{\partial}{\partial z}\left( {\rm sign}(z)\, f\, V_{\rm conv} \right) = Q-2h\, \delta(z)\, \Gamma_{\rm ann} f  ,
\eeq
 acquires a simple factorized form in the ``no-tertiaries" approximation that we adopt.
 The $\bar d$ flux in the galactic medium around the solar system can be written as
\beq
\frac{d\Phi_{\bar d}}{dT} =\frac{v_{\bar d}}{4\pi} f =\frac{v_{\bar d}}{4\pi}  \frac{ \langle \sigma v\rangle}{2} \left(\frac{\rho_\odot}{M}\right)^2 R_d(T)\frac{dN_{\bar d}}{dT},
\label{eq:RT}
\eeq
fully analogous to the solution for the $\bar p$ flux in~\cite{MDM3}.
The function $dN_{\bar d}/dT$ contains the particle physics input and was computed in the previous section.
The function $R_d(T)$ encodes the Milky Way
astrophysics and is plotted in fig.\fig{HaloI}b for various halo and propagation models.
It roughly is some average containment time in the diffusion cylinder, and we verified that $\bar d$ generated outside
it provide a negligible extra contribution even in the min scenario, where 
most DM annihilations occur outside the diffusion cylinder: the probability of re-entering is sizable, but
the probability of diffusing up to the solar system is small.
Going from DM annihilations to DM decays with life-time $\tau$ one just needs to replace in eq.\eq{RT}
$\langle \sigma v\rangle \rho_\odot^2/2M^2$ with
$\rho_\odot/M\tau$; we do not plot the corresponding $R_d(T)$ functions for DM decay as they essentially coincide
with the $R_d$ function for DM annihilations and the {\em isothermal} profile plotted in fig.\fig{HaloI}b.
Indeed, for all the considered DM profiles, DM decays close to the galactic center do not significantly contribute 
to the $\bar d$ flux at Earth, as for DM annihilations with  the quasi-constant isothermal density profile.

We notice that although $R_d(T)$ is significantly uncertain (especially below a few GeV), 
the ratio with the corresponding astrophysical function $R(T)$ for $\bar p$ is
essentially fixed, so that the non-observation of a DM $\bar p$ excess puts robust bounds on the possible
DM $\bar d$ flux.
Indeed the $\bar p$ flux has been observed below 100 GeV by PAMELA~\cite{PAMELApbar} and agrees with astrophysical expectations,
which are believed to have an uncertainty of about $\pm 20\%$~\cite{Fornengo,Fornengo20}.

Finally, we take into account the solar modulation effect, relevant only for non-relativistic $\bar d$: the solar wind decreases the kinetic energy $T$ 
of charged cosmic rays such that the energy spectrum $d\Phi_{\bar{d}\oplus}/dT_\oplus$
of $\bar d$ that reach the Earth with energy $T_\oplus$ 
 is approximatively related to their energy spectrum in the interstellar medium, $d\Phi_{\bar d}/dT$,
as~\cite{GA}
\beq \frac{d\Phi_{{\bar d}\oplus}}{dT_\oplus} = \frac{2m_d T_\oplus + T_\oplus^2}{2 m_d T + T^2} \frac{d\Phi_{\bar d}}{dT},\qquad
T= T_\oplus + e \phi_F.\eeq
The so called Fisk potential $\phi_F$ parameterizes in this effective formalism the kinetic energy loss. 
We assume $\phi_F= 0.5\, {\rm  GV}$ i.e.\
$e\phi_F= 0.5\, \GeV$.

\begin{figure}
\begin{center}
\includegraphics[width=0.95\textwidth]{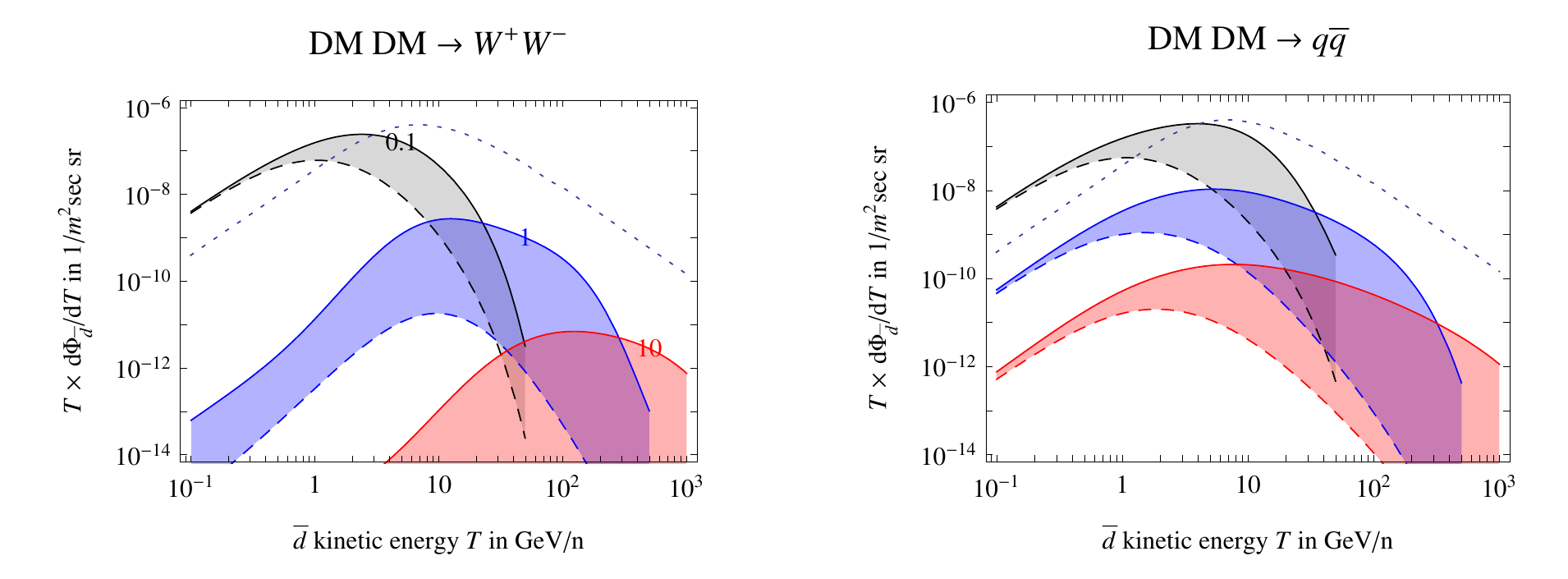}
\caption{\em\label{fig:fluxescompare} Assuming the $\sigma v = 3\cdot 10^{-26}\cm^3/{\rm sec}$ suggested
by cosmology, the NFW profile, MED propagation
and DM masses $M=\{0.1,1,10\}\TeV$, we compare the $\bar d$ flux obtained from
the full computation (continuous lines) with the one from the spherical approximation.
The dotted line is the expected astrophysical background.
}
\end{center}
\end{figure}

\begin{figure}
\begin{center}
\includegraphics[width=0.95\textwidth]{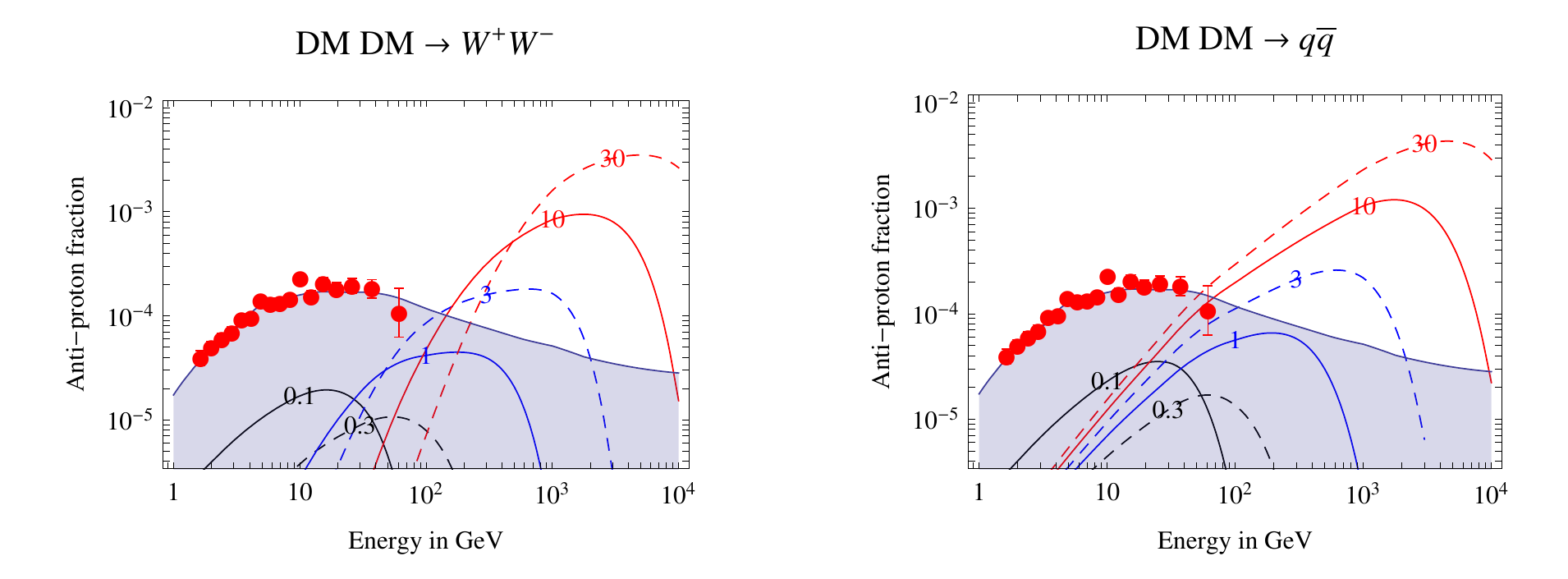}
\caption{\em\label{fig:fluxesp} 
{The $\bar p/p$ ratio}, for the
same DM models described in the caption of fig.\fig{fluxes}, showing that they are
compatible with the PAMELA $\bar p$ data (red points).
Shading indicates the expected astrophysical background.}
\includegraphics[width=0.95\textwidth]{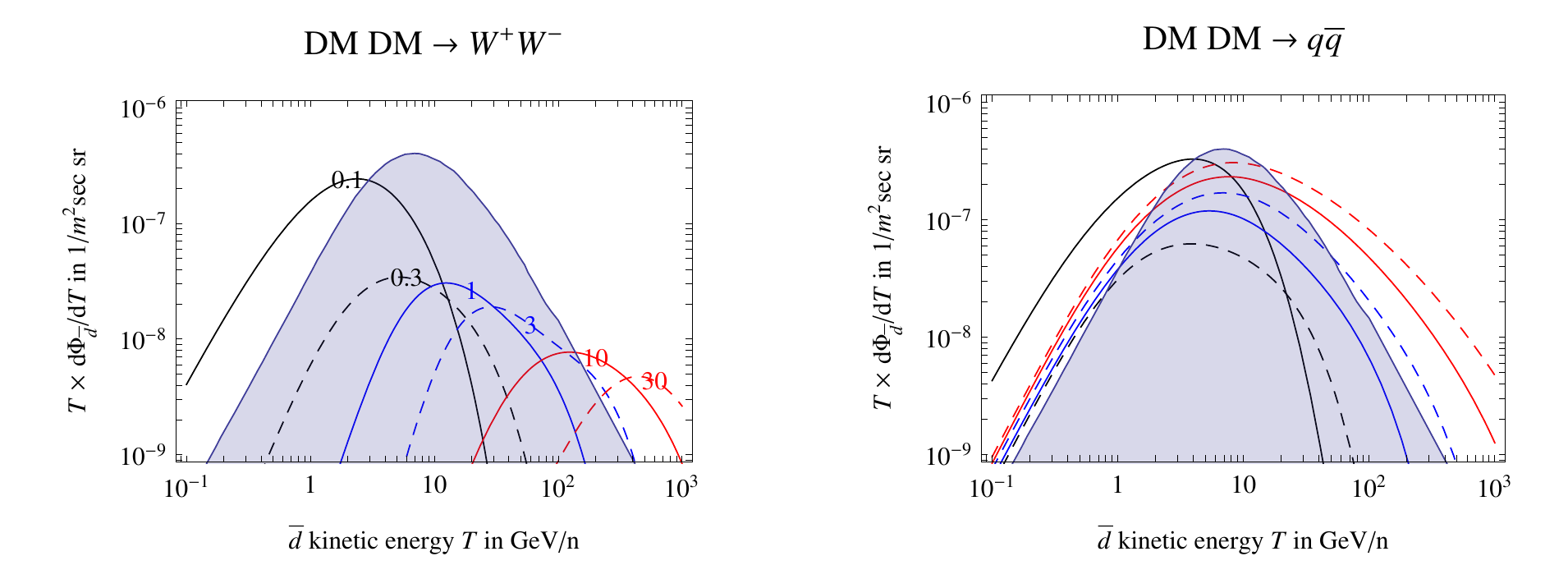}
\includegraphics[width=0.95\textwidth]{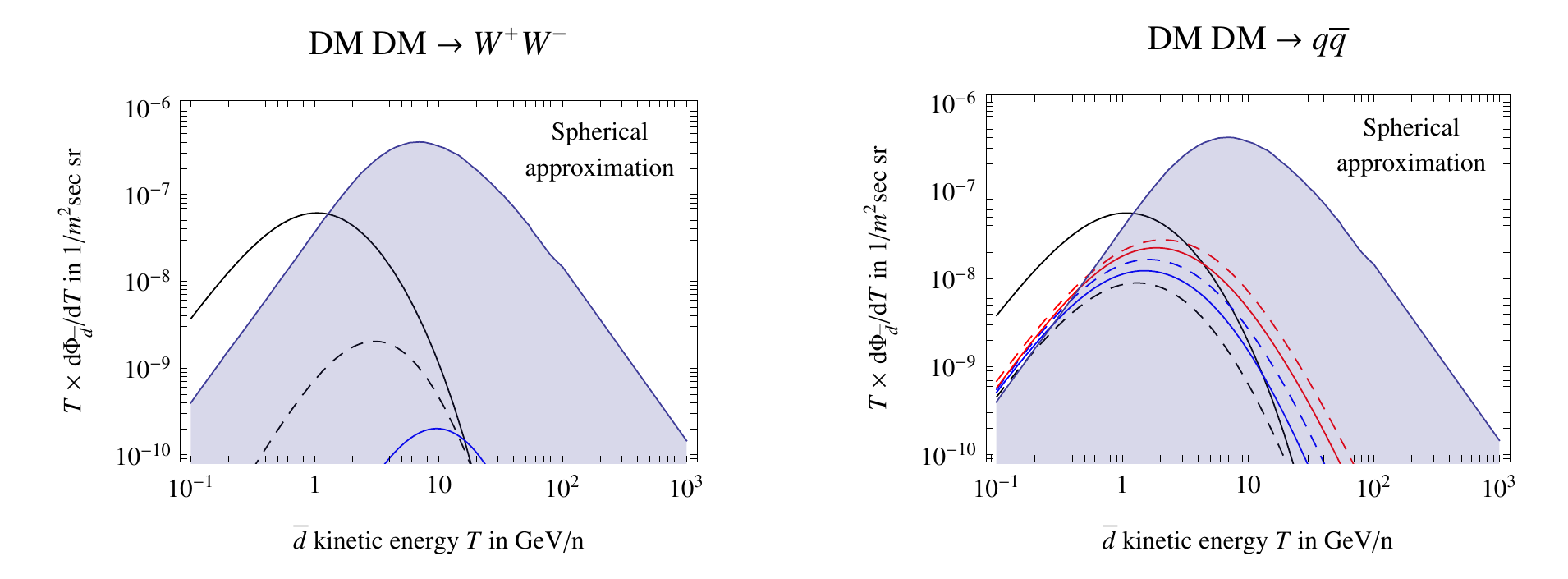}
\caption{\em\label{fig:fluxes} Upper row:
our result for the $\bar d$ flux at Earth.
Lower row: previous results for the $\bar d$ flux computed in the spherical approximation.
We consider DM masses $M=0.1,1,10\TeV$ (black, blue, red continuous curves)
and $M=0.3,3,30\TeV$ (black, blue, red dashed curves),
DM annihilations into $W^+W^-$ (left) and $q\bar q$ (right) with
$\sigma v = 3\cdot 10^{-26}\cm^3/{\rm sec} \times \max(1,M/300\GeV)^2$, the NFW DM profile, MED propagation,
solar modulation $\phi_F =0.5\,{\rm GV}$, $p_0 = 160\MeV$.
Shading indicates the expected astrophysical $\bar d$ background.
}
\end{center}
\end{figure}

\section{Results}
Fig.\fig{fluxescompare} compares our Monte Carlo results for the $\bar d$ flux with the spherical approximation.\footnote{
Numerical results in some previous computations apparently included spurious factors of 2
related to $dN_p/dx ({\rm after~neutron~decay)}\approx 2 dN_p/dx ({\rm before~neutron~decay)}$ 
(this explains a discrepancy with~\cite{Fornengo}) 
and to GeV/nuc = GeV/2  (that affects the measure $dT$ in $d\Phi_{\bar d}/d T$;
when comparing our plots with ones in previous papers, 
notice that we plot $d\Phi_{\bar d}/d\ln T$ rather than $d\Phi_{\bar d}/d T$).
}
The shading indicates the enhancement.
We here assumed the NFW profile, MED propagation. and the  DM annihilation cross section
$\sigma v=\sigma v_{\rm cosmo} \equiv 3~10^{-26}\cm^3/{\rm sec} $
that reproduces the cosmological DM abundance via thermal freeze-out.

Rather than relying on theoretical assumptions, in order to explore the maximal $\bar d$ flux from DM
compatible with present data, we assume
\beq \sigma v = \max (1, M/300\GeV)^2\cdot \sigma v_{\rm cosmo}\  .\eeq
Indeed, fig.\fig{fluxesp} shows that these assumptions give a 
$\bar p$ flux compatible with (and comparable to) the PAMELA $\bar p/p$ data.
As discussed in the previous section, the $\bar p/\bar d$ ratio is negligibly affected by astrophysical uncertainties.
Furthermore, the assumed cross section 
is about one order of magnitude below what is needed to explain  the PAMELA~\cite{PAMELA} $e^+$ excess 
and is compatible with the bounds  from galactic $\gamma$ and $\nu$ observations~\cite{bounds} as well as with
the diffused $\gamma$-ray constraints \cite{Huetsi:2009ex}.


Then, the upper row of fig.\fig{fluxes} shows our Monte Carlo results for the $\bar d$ flux, $T\cdot d\Phi_{\bar d}/dT,$ while
the lower row shows the corresponding much lower $\bar d$ flux obtained in the spherical approximation.
The caption describes all various assumptions.

Comparison of these two results shows that the signal is 
 enhanced in our Monte Carlo $\bar d$ computation:
 almost an order of magnitude for 
 small $\bar d$ kinetic energies ($T\sim1$~GeV) or lighter DM ($M\sim 100\GeV$)
 and orders of magnitude at higher energies or for heavier DM.
 Such enhancement does not depend on the assumed value of $\sigma v$
 Before our calculation it was believed that only the sub-GeV energy region is suitable for searches of a DM-induced 
 $\bar d$ signal.  Our result implies that heavy DM, as suggested by PAMELA and FERMI data, 
 also induces detectable $\bar d$ signal at high energies.
Therefore our result has important implications on the strategy of DM searches using the  $\bar d$ signal.


The red line in fig.\fig{fluxes}a
roughly shows the Minimal Dark Matter~\cite{MDM3} prediction:
DM with $M\approx 10\TeV$ that makes Sommerfeld-enhanced annihilations into $W^+W^-$,
giving rise to $\bar p$ and consequently to $\bar d$ at energies (per nucleon) above $m_p M/M_W$,
not yet explored by PAMELA.

The PAMELA~\cite{PAMELA}, FERMI~\cite{FERMI}
 and HESS~\cite{HESSepm} $e^\pm$ excesses suggest a DM interpretation in terms of multi-TeV
DM that annihilates dominantly into leptons with a Sommerfeld-enhanced cross section~\cite{CKRS, Somm}.
An interesting class of models with these properties
is obtained by assuming that DM annihilates into a new vector 
with mass $m<2m_p$, that subsequently can only decay into the lighter 
$e,\mu,\pi$~\cite{Nima}.  We notice that this condition is not strictly necessary neither for the Sommerfeld enhancement
nor for compatibility with
PAMELA $\bar p$ data: indeed if  $m\circa{>}2m_p$ one would obtain $\bar p$ 
with energy larger than $m_p M/m$, where $M/m$ is the boost factor of the new vector.
This boost is large enough not to give an unseen $\bar p$ excess below 100 GeV (the energy range explored by PAMELA so far)
even if $m$ is several tens of GeV, as in~\cite{CKRS}.
Similarly to the Minimal Dark Matter case, these models would give a flux of $\bar d$ above 100 GeV.
Our enhanced $\bar d$ signal should also be used to re-evaluate prospects of
discovering supersymmetric Dark Matter
candidates, which often annihilate into the $W^+W^-$ or $b\bar b$ modes we considered.

\section{Conclusions}

We computed the  $\bar d$ flux at Earth produced by   DM annihilations or decays in the Milky Way
using an  event-by-event Monte Carlo technique run on the {\sc Grid}, improving on previous computations
that assumed spherically symmetric events and obtained a 
 $1/M^2$ suppression of the  $\bar d$ yield for heavy DM masses $M$.
Due to the
jet structure of high energy events implied by relativity
no such suppression is present,
and the $\bar d$ signal is strongly enhanced: by orders of magnitude for $\bar d$ energies above 10 GeV
or DM masses above 1 TeV, as illustrated in fig.\fig{fluxes}.
The $\bar d$ astrophysical background seems not to be significantly affected, 
being dominantly generated by low-energy cosmic ray collisions.
While the $\bar p$ and $\bar d$ fluxes suffer from significant astrophysical uncertainties,
their ratio is robustly predicted.  Thereby the non-observation of a $\bar p$ excess in PAMELA data
implies an upper bound on the $\bar d$ DM flux.
In the light of our enhanced $\bar d$ fluxes, we find that a $\bar d$ DM signal is still possible.
For example, heavy DM models~\cite{MDM3,Nima} that can account for the PAMELA $e^\pm$ excess
can lead to  $\bar p$ and $\bar d$ excesses above 100 GeV/nucleon.

Most importantly, our result implies that the experiments searching for cosmic ray $\bar d$
become sensitive to $M\geq$~TeV mass DM,
provided that the DM annihilation cross section is larger than what naively suggested
by thermal freeze-out.
Therefore it is important to extend future searches for $\bar d$  above the GeV energy range.
For the moment, the AMS-2 experiment
is expected to achieve a very energy-dependent efficiency to $\bar d$ detection,
so that AMS-2 would have a sensitivity to a $\bar d$ flux down to
$5~10^{-7}/({\rm m}^2\sec\sr\GeVn)$ in the energy ranges
$0.2\GeVn<T<1\GeVn$ (where time-of-flight is enough to discriminate $\bar d$ from $\bar p$)
and $2\GeVn<T<4\GeVn$ (where the magnetic spectrometer is needed)~\cite{AMS}.
According to previous $\bar d$ DM computations based on the spherically symmetric approximation, 
only the lower energy range was promising for DM searches.  
We have shown that the DM signal can manifest itself also at higher energies, where it is 
less affected by astrophysical uncertainties.

\paragraph{Acknowledgements.} 
We thank Nicolao Fornengo, Ignazio Bombaci, Alejandro Kievsky, Torbj\"orn Sj\"ostrand,
Antonello Polosa, Michele Viviani
and especially Marco Cirelli for useful conversations.
This work was supported by the ESF Grant 8090, Estonian Ministry of Education and Research project SF0690030s09 and by EU FP7-INFRA-2007-1.2.3 contract No 223807.

\appendix

\footnotesize

\begin{multicols}{2}

\end{multicols}


\begin{thebibliography}{nn}

\bibitem{Csernai} L.~P.~Csernai and J.~I.~Kapusta,
  Phys.\ Rept.\  {131} (1986) 223.

\bibitem{d-bar}
F.~Donato, N.~Fornengo and P.~Salati,
  Phys.\ Rev.\  D {62} (2000) 043003
  [arXiv:hep-ph/9904481].


\bibitem{d-bar2}
  H.~Baer and S.~Profumo,
  JCAP {0512} (2005) 008
  [arXiv:astro-ph/0510722].


\bibitem{Fornengo} 
 \hepart[0803.2640]{F. Donato, N. Fornengo, D. Maurin}.
 See also \art[hep-ph/0503544]{R. Duperray et al.}{\PR}{D71}{083013}{2005}.
 N. Fornengo, private communication.
 

 \bibitem{Fornengo20}
 \hepart[0909.4548]{G. Di Bernardo, C. Evoli, D. Gaggero, D. Grasso, L. Maggione}.


\bibitem{Cirelli} \hepart[0904.1165]{C.B. Br\"auninger, M. Cirelli}.


\bibitem{Ibarra} \hepart[0904.1410]{A. Ibarra, D. Tran}.
  


\bibitem{cow} \href{http://en.wikipedia.org/wiki/Spherical_cow}{SphericalCow at wikipedia}.


\bibitem{ALEPH} \art[hep-ex/0604023]{ALEPH collaboration}{Phys. Lett.}{B369}{192}{2006}.


\bibitem{Pythia}
  T.~Sjostrand, S.~Mrenna, P.~Skands,
  Comput.\ Phys.\ Com.\  {178}, 852 (2008)
  [arXiv:0710.3820].

\bibitem{BG}
\href{http://www.balticgrid.org}{www.balticgrid.org}.

%


\bibitem{rho(r)}
Isothermal profile:
\art{J. N. Bahcall and R. M. Soneira}{Astrophys. J. Suppl.}{44}{73}{1980}.\\
NFW profile:
\art[astro-ph/9611107]{J. Navarro, C. Frenk, S. White}{Astrophys. J.}{490}{493}{1997}.\\
Einasto profile: 
J. Einasto, 
1965, Trudy Astrophys. Inst. Alma-Ata, 5, 87; Tartu Astron. Obs. Teated Nr. 17.
See also:
  \hepart[0810.1522]{J.~F.~Navarro {\it et al.}}.\\
Moore profile:
\art[astro-ph/0402267]{J.~Diemand, B.~Moore and J.~Stadel}{Mon.\ Not.\ Roy.\ Astron.\ Soc.}{353}{624}{2004}.


 


\bibitem{minmedmax}
  F.~Donato, N.~Fornengo, D.~Maurin and P.~Salati,
  Phys.\ Rev.\  D {69}, 063501 (2004)
  [arXiv:astro-ph/0306207].
  
  


\bibitem{MDM3}   
\art[0802.3378]{M.~Cirelli, R.~Franceschini, A.~Strumia}{Nucl.\ Phys.}{B800}{204}{2008}.


\bibitem{PAMELApbar}
\hepart[0810.4994]{PAMELA collaboration}.

 


\bibitem{Kane} \hepart[0906.4765]{G. Kane et al.}.


\bibitem{CKRS}
\art[0809.2409]{M. Cirelli, M. Kadastik, M. Raidal, A. Strumia}{\NP}{B813}{308}{2009}.


\bibitem{GA}
\art{L.J. Gleeson and W.I. Axford}{ApJ}{154}{1011}{1968}.


\bibitem{PAMELA}
\hepart[0810.4995]{PAMELA collaboration}.


\bibitem{bounds}
 \art[0811.3744]{G. Bertone, M. Cirelli, A. Strumia, M. Taoso}{JCAP}{0901}{2009}{43}.
J.~Hisano, M.~Kawasaki, K.~Kohri and K.~Nakayama,
  arXiv:0812.0219 [hep-ph].
  \hepart[0905.0480]{P. Meade, M. Papucci, A. Strumia, T. Volansky}.
\hepart[0905.2075]{J.~Hisano, K.~Nakayama and M.~J.~S.~Yang}.


\bibitem{Huetsi:2009ex}
  G.~Huetsi, A.~Hektor and M.~Raidal,
  arXiv:0906.4550 [astro-ph.CO].


\bibitem{FERMI} \hepart[0905.0025]{FERMI/LAT collaboration}.


\bibitem{HESSepm} 
\hepart[0811.3894]{H.E.S.S. collaboration}.\\
\hepart[0905.0105]{H.E.S.S. collaboration}.


\bibitem{Somm}
A. Sommerfeld, Ann. Phys. 11 257 (1931).
J.~Hisano, S.~Matsumoto and M.~M.~Nojiri,  Phys.\ Rev.\ Lett.\  {92} (2004) 031303  [hep-ph/0307216].
\art[0706.4071]{M.~Cirelli, A.~Strumia, M.~Tamburini}{Nucl.\ Phys.}{B787}{152}{2007}.


\bibitem{Nima} N.~Arkani-Hamed, D.~P.~Finkbeiner, T.~Slatyer and N.~Weiner,
arXiv:0810.0713.  


\bibitem{AMS}
V. Choutko and F. Giovacchini, {\em ``Cosmic rays $\bar d$ sensitivity for AMS-02 experiment''},
talk at the ICRC07 conference.


\end{thebibliography}
\end{document}